\documentstyle[spie]{article}
\input{psfig}
\title{Imaging performance of the XMM-Newton X-ray telecopes}

\author{B. Aschenbach\supit{a}, U. Briel\supit{a}, F. Haberl\supit{a},
H. Br\"auninger\supit{a}, W. Burkert\supit{a}, A. Oppitz\supit{a},\\ 
P. Gondoin\supit{b} and D. Lumb\supit{b}
\skiplinehalf
\supit{a}Max-Planck-Institut f\"ur extraterrestrische Physik, \\
Giessenbachstrasse, 85740 Garching, Germany
\\
\supit{b}European Space Research and Technology Center, \\
2200 AG Noordwijk, The Netherlands
}

\begin{document}
  \maketitle
\begin{abstract}
The in-orbit imaging performance of the three X-ray telescopes on board 
of the X-ray astronomy observatory XMM-Newton is presented and compared 
with the performance measured on ground at the MPE PANTER test 
facility. The comparison shows an excellent agreement between 
the on ground and in-orbit performance. 
\end{abstract}

\section{INTRODUCTION}

On December 10 1999 the XMM spacecraft was placed in a 48h Earth orbit 
by the first commercial ARIANE V launcher. 
XMM, or X-ray Multi-Mirror Mission, is the second cornerstone of the 
Horizon 2000 science program of the European Space Agency ESA. 
XMM was designed as a  facility class X-ray astronomy observatory to study 
cosmic X-ray sources spectroscopically with the highest collecting area 
feasible in the 0.1 - 10 keV band. 
Over the full band moderate spectral resolving power between 1 and 60 
is required and medium resolving power of $\ge$ 250 was aimed for 
in the 0.1 - 3 keV band. The high throughput is primarily achieved by a set 
of 3 imaging, highly nested Wolter type I telescopes. The imaging performance 
of each of the grazing incidence mirror systems was originally set to 
 be better than 
30 arcsec for the half energy width (HEW) of the point spread function 
(PSF) with a
 goal of 10 arcsec. 
In a first attempt the scientific requirements have been put together at a 
workshop 
held at Lyngby, Denmark, in June 1985$\sp 1$. A detailed analysis of the 
types
of telescopes with which these requirements could be achieved was carried out 
by the ESA established Telescope Working Group, which also arrived at a 
first telescope design$\sp{2,3}$. Because of launcher volume constraints 
which became apparent shortly after the publication of the Telescope 
Working Report  the 
baseline design was slightly modified in March 1987$\sp 4$. Since then 
the design was not changed any more and the actual 
telescopes and their mirrors, now being in orbit, are of that design.
Over ten years in time the design was put into reality, the telescopes 
were built in industry including Carl Zeiss, Oberkochen in Germany, 
Kayser-Threde, Munich in Germany and Media Lario, Bosisio Parini in Italy. 
The development and building of the mirror shells up to the flight mirror 
modules was accompanied by X-ray tests 
in the 130 m long X-ray test facility PANTER of the Max-Planck-Institut 
f\"ur extaterrestrische Physik$\sp 5$. The qualification of the mirror 
modules was performed in EUV light at the Focal-X facility of the 
Centre Spatial de Li\`ege in Belgium.

After a short description of the telescope design and its critera followed 
by an outline of the manufacturing process of the mirrors the in-orbit imaging 
performance of each of the three X-ray telescopes is presented and 
discussed and confronted with the full X-ray beam calibration 
measurements taken at the PANTER.

\section{OPTICS DESIGN}

The design of the optics was driven by the requirement of obtaining the 
highest possible effective collecting area over a wide band of energies up 
to 10 keV and more with particular emphasis on the region around 
7 keV, in which the K lines of the astrophysically important iron appear. 
Thus, the mirror system has to utilize a very shallow grazing 
angle of about half a degree in order to provide sufficient reflectivity 
at high energies. Both, Wolter type and Kirkpatrick-Baez type telescopes 
can in principle be considered for the optics. The Kirkpatrick-Baez system 
has however a focal ratio, defined by the ratio of the focal length over the 
diameter of the optics, which is two times greater than that for a corresponding 
Wolter type I system. Taking the same focal length a Wolter type I system 
can therefore make efficient use of a mirror system at the 
same grazing angle which is a factor of two wider and an increase in 
collecting area by a factor of 
four. This factor of four is slightly reduced to a factor of $\pi$ since the 
Kirkpatrick-Baez system builds rectangular whereas the Wolter system 
has circular mirrors. Because of this major advantage the Telescope 
Working Group baselined the Wolter type I system for XMM.

In grazing incidence optics the effective area is generally increased by 
nesting a number of mirrors and thus filling the front aperture as far 
as possible. The 
nesting efficiency is determined by the mirror shell thickness and, 
in case of very low grazing angles for high energy optics, by the 
minimum radial mirror separation which is required for integration 
and alignment. In soft X-ray telescopes of moderate grazing angles 
the separation is of less importance since adjacent shells are 
ususally comfortably separated anyway to prevent blocking of on-axis 
rays. 
 
The thinner the mirror shells are and the narrower the shells are spaced the 
larger is the collecting area. In the past very thin aluminum foils shaped to  
double cones have 
been used, for instance for the BBXRT telescopes, the telescopes on ASCA and 
ASTRO-E. Due to the usage of foils and the very narrow foil 
placement this type of telescope has the potential of achieving 
maximal collecting area. There are, however, two major drawbacks which 
are inherent to foil telescopes. Firstly, the mirror sections are bent into 
cones, which produce an extended image even for 
on-axis point-like sources, leaving significantly less tolerances for 
mirror surface irregularities and mirror mounting. Secondly, foils have 
been reported to have intrinsic mid-frequency irregularities which limit 
the angular resolution to something like one arcmin and more. Because of these 
disadvantages foil telescopes were considered by the Telescope Working 
Group not to meet the XMM resolution requirements. 

Thin-walled mirrors have also been produced by replication techniques 
and electroforming. Electroformed mirrors have been built up mostly in 
solid nickel, which, however, suffer from too much mass, and they were 
therefore initially discarded by the Telescope Working Group. Development 
work on the mirrors started by producing carbon fibre reinforced 
plastics (CFRP) mirrors by epoxy replication from master mandrels 
polished to X-ray imaging quality. 
Advantages of this approach are mirrors of high stiffness with thin walls, 
implying low mass, and low thermal expansion. Disadvantages are the 
non-amorphous structure of the CFRP implying the development of 
unacceptable surface structure (print-through) and sensitivity to 
humidity (out-gassing or even "breathing") thereby changing the surface 
with environment.

At the time of development start CFRP mirror carriers could be made as 
thin as 0.5 mm for the smallest diameter shells, which is the 
thickness baselined for the 
optics design calculations. The shell thickness is assumed to increase 
approximately linear with shell diameter to guarantee sufficient stiffness. 
A minimum radial shell separation of 1 mm was considered to be feasible 
to handle the integration of the shells in a package. Both the paraboloid 
section and 
the hyperboloid section were replicated as a single piece from a 
single mandrel, which limits the axial length of the total mirror to 60 cm. 
The length  
is shared evenly by the paraboloid and the hyperboloid. Although 
 longer mirrors would have provided a larger collecting area they were 
discarded because they  
were suspected to be very difficult to be removed from the mandrel. 
The production of paraboloid and hyperboloid in one piece avoids the 
problems created by the additional tasks of aligning and mounting 
if paraboloid and hyperboloid are separate. Given the length constraints 
imposed by the launcher the focal length was fixed at 7.5 m. The optimum 
design, which fulfils the collecting area constraints, was found by 
ray tracing. Gold was considered as baseline for the mirror surface 
coating; attempts to produce replicated mirrors with an iridium coating, 
which would have provided somewhat better high energy response, failed 
because the adhesive forces of iridium on the mandrel surface turned 
out to be too strong so that patches of the iridum layer tended to 
stick on the mandrel.     
The parameters of the optimum design are shown in Table~\ref{tab:design}. 
The mirror 
diameter of the outermost shell is 700 mm. At low energies $<$ 2 keV 
the area could have been increased by adding larger shells. In contrast, 
the area at 7 keV increases up a mirror diameter of 650 mm but drops for 
larger diameter shells. Adding more shells to the inner region would increase 
the higher energy response but at rather low efficiency, which means 
that the area per shell is significantly less than the area averaged 
over all shells. The optimum telescope design is therefore defined by a 
maximum 
diameter mirror shell of 700 mm and 58 mirror shells in total. It is 
optimal because it has  
 a 7 keV response which is only marginally lower than maximal. On top 
it has  very good low energy response. 

\begin{table} [h]
\caption{XMM-Newton Wolter Type I X-ray telescope characteristics.}
\label{tab:design}
\begin{center}
\begin{tabular}{|l|r|}
\hline
\rule[-1ex]{0pt}{3.5ex}Focal length &                         7500 mm \\
\rule[-1ex]{0pt}{3.5ex}
Outer mirror radius    &                350.0 mm \\
\rule[-1ex]{0pt}{3.5ex}
Inner mirror radius     &               153.1 mm  \\
\rule[-1ex]{0pt}{3.5ex}
Axial mirror length (par. + hyp.) &     600 mm   \\
\rule[-1ex]{0pt}{3.5ex}
Outer mirror wall thickness        &    1.07 mm  \\
\rule[-1ex]{0pt}{3.5ex}
Inner mirror wall thickness         &   0.47 mm  \\
\rule[-1ex]{0pt}{3.5ex}
Minimum nesting distance             &  1 mm  \\
\rule[-1ex]{0pt}{3.5ex}
Reflective surface material         &   gold  \\
\rule[-1ex]{0pt}{3.5ex}
Number of mirrors per telescope      &  58    \\
\rule[-1ex]{0pt}{3.5ex}
Mirror surface per telescope        &   51.4 m$\sp 2$ \\
\rule[-1ex]{0pt}{3.5ex}
Number of telescopes                 &  3    \\
\hline 
\end{tabular}
\end{center}
\end{table}

After several years of study, development, building and X-ray testing of 
CFRP mirrors the angular resolution requirement of 30 arcsec could not be 
met. Mirrors of $<$ 1 arcmin were produced several times but their 
susceptibility to humidity appeared to be an unsurmountable problem. 
Inspired by the progress which was made in the production of good 
electroformed nickel shells in the context of the SAX and JET-X programs, 
it was decided in 1991/92 by the ESA XMM Project Team in consultation with the 
 XMM Telescope Advisory Group to terminate the CFRP activities and 
to build the XMM mirrors from electroformed nickel.    
The original telescope design was kept except that the mirror 
shells were reduced in wall thickness by about 25\% \ to save mass 
as much as possible 
 (Table~\ref{tab:design} shows the mirror wall thicknesses for the nickel 
mirrors). Because of the high mass involved with nickel made 
mirror shells the number of  X-ray telescopes was definitely limited to 
three. The high mass also contributed to  the discussion of switching 
launchers  from ARIANE IV to ARIANE V.  

The production of nickel shells involves three steps. As for any replication 
a master mandrel of negative shape is produced for each of the 58 mirror 
shells. The mandrel is of solid aluminum covered with a thin layer of 
Kanigen, which is ground and polished to the precise negative shape 
of the required mirror. The mandrel surface is measured by  
metrology means, including profilometers of various trace lengths and 
interferometers. Ray tracing of the metrology data reveals that the 
XMM mandrels have an angular resolution of $\le$ 5 arcsec HEW  
and an rms surface microroughness of $\le$ 0.5 nm. 
The mandrels are coated with a gold layer about 200 nm thick. 
Each mandrel is put in an electrolytical bath in which a nickel layer 
is being built upon the gold surface until the required thickness has 
been reached. By cooling the nickel shell  separates from the 
mandrel, whereby the gold surface sticks to the electroproduced 
nickel shell because of the higher adhesion forces.
After separation the nickel shells are exclusively supported by 16 strings 
threaded through 16 tiny holes which are equally spaced around the 
circumference of the mirror closely below the rim of the paraboloid.  
After inspection of the  surface and measurement of  a couple of meridional 
profiles each mirror is mounted to a spider like support structure by 
glueing the parabola mirror into small grooves in each of the 16 arms of 
 the spider. This is done with the mirror 
oriented vertically and under optical control. 
The spider support structure at the paraboloid front is the only support 
structure for the entire mirror module. There is no counterpart at the 
aft of the hyperboloids; they are not constrained or supported mechanically 
at all. 
Further details about the production of the XMM telescopes are given in  Ref. 6.

From an X-ray optical point of view the telescope is complemented by an 
X-ray baffle, which does not 
eliminate but significantly 
reduces X-ray straylight from sources outside the field of 
view. Sources outside the field of view can reach the sensitive 
area of the focal plane detectors by single reflection from the 
rear end of the hyperbola. Less highly nested telescopes like 
those on the Einstein, ROSAT and Chandra observatories employ 
radial vanes with one set inside and one set in front of 
each mirror to eliminate singly reflected rays. 
For XMM an external multi-stage baffle has been conceived$\sp 7$, 
which basically consists of a set of  concentric cylinders, one 
for each mirror shell, placed 
in front of the mirror system. Each ring has an annular width slightly 
smaller than the corresponding mirror wall thickness and a cylindrical height 
characteristic for each mirror by which the singly reflected rays are 
blocked. 
The implementation of such a system would have eliminated the straylight 
completely. But because of the mechanical complexity together with the 
limited space available 
the design 
was simplified to two parallel annular plane  sieve plates of equal thickness. 
Each plate consists of 58 rings and 16 radial struts. The plates are mounted 
co-axial to and co-aligned 
with the front aperture cross section of the 58 mirror shells. 
The off-set  of the two sieve plates from the front of the mirror 
system is 385 mm and 439 mm, respectively$\sp 8$. 
The efficiency of the sieve plate system has been ray traced 
 and demonstrated to reduce 
the straylight level by a factor of 5 to 10 depending on the 
position in the focal plane.

A complete overview of the telescope with all its subsystems 
including the mirror door, the entrance optical baffle, 
the electron deflector and additional support structure is given in 
Ref. 6.

\section{X-RAY CALIBRATION ON GROUND}

During Phase C/D of the XMM program five X-ray mirror modules have been built
three of which have been selected for flight, i.e. FM2, FM3 and FM4. FM1 
and FM5 have been kept as spares, numbering of the modules is in 
chronological order of production. 
X-ray testing and final X-ray calibration including measurement of the
PSF both on-axis and off-axis for a few energies 
between 0.28 keV and 8.05 keV and the effective area as well as the  vignetting 
function for a series of energies covering the range from 0.28 up to 
17 keV has been performed in MPE's PANTER test facility between 
April 1997 (FM1) and July 1998 (FM4). FM5 was tested in 1999 but it has not 
been completed to the extent to serve as flight spare, whereas FM1 actually 
is fully functional for flight.  
MPE's PANTER originally built for the development 
and calibration of the ROSAT X-ray telescope was modified and partially 
rebuilt in 1992/93  to cope with the size of the XMM telescopes 
and quality assurance requirements. In simple terms, 
the facility consists of a 130 m long vacuum pipe of 1 m diameter. 
At the one end a micro-focus characteristic X-ray line source is installed 
and the use of various anti-cathode materials provides the required 
coverage of the energy band. The bremsstrahlung component can be used as 
a continuum source as well. The X-ray source has a diameter of about 1 mm. 
At the other 
end a huge cylindrical vacuum 
 tank 12 m long and 3.5 m in diameter houses the X-ray telescope to be tested. 
The typical distance between 
source and mirror module midplane, i.e. the paraboloid hyperboloid 
intersection, is about 124 m. With a focal length of 7.5 m the image of 
the X-ray 
source is produced at a distance of 7.97 m measured from the mirror 
module midplane. Given these distances the angular size of the X-ray 
source is about 1 arcsec, which means that structures of the mirror 
surface finer than this cannot be resolved.

Here we report about the measurements with the entire 
aperture of the mirror module illuminated, implying a beam divergence 
between 8.5 arcmin and 19.5 arcmin full width between 
the innermost and outermost mirror shells. The divergence of the 
illuminating beam implies that the rays reflected off the frontal one third of 
the parabolas are not reflected from the hyperbolas. Therefore the 
surface of this section of the parabolas is not imaged and does not 
contribute to the PSF. Furthermore the corresponding 
reflection loci on parabola and hyperbola differ from those for 
parallel beam illumination. 
Images have been recorded with both  a position sensitive 
proportional counter (PSPC) which is a copy of the PSPCs flown on-board 
of ROSAT$\sp 9$ and an X-ray CCD provided by the X-ray astronomy group 
of the University of Leicester in the UK. Whereas the PSPC is well suited to 
measure effective area and the wings of the telescope point spread function 
the CCD with it significantly better spatial resolution has been used 
to assess the details of the PSF out to angular 
distances of about 125 arcsec limited by the linear size of the CCD. 
The PSPC has a larger field of view and the PSF can be traced 
out to angular distances of about 1000 arcsec. The comparision of the 
CCD and the PSPC encircled energy functions show that the CCD images 
out to 125 arcsec contain 97\% \ of the total flux at 1.5 keV and 
93\% \ at 8.05 keV, respectively.
The results have been reported in Ref. 10,11,12 and a summary 
is given in Ref. 6.

\section{THE POINT SPREAD FUNCTION IN ORBIT}

On January 19 2000 the X-ray telescope FM2 saw "First Light" of 
the X-ray sky$\sp{13}$ followed by FM3 and FM4 two days later. 
Two different X-ray cameras are used to record the X-ray images. 
In the focal plane of FM2 a novel type of X-ray CCD based on pn-technology 
is used (FM2/pn)$\sp{14}$ whereas FM3 and FM4 each are equipped with 
a more conventional CCD based on MOS-technology 
(FM3/MOS1 and FM4/MOS2). Unlike FM2, FM3 and FM4 each carry 
a reflection grating spectrometer (RGS)$\sp{15}$ by which about 
half of the X-ray beam is diverted from MOS1 and MOS2.  
At a press conference on February 9 2000 the "First Light" results 
were presented and the new name of the mission XMM-Newton was announced.
After "First Light" the imaging performance of the three telescopes was
of immediate interest and additional observations were made during the 
commissioning phase of XMM-Newton until March 8. Sources studied 
include the sources of the "First Light Field" to the Large Magellanic Cloud 
and 
the point sources EXO 0748-67, LMC X-3, PSR 0540-69 and PKS 0558-504. 
A first analysis shows that the observations,
 except for LMC X-3 at high energies, taken in full frame mode of the CCDs, 
i.e. the entire field of view is read out from the CCD at once, are 
not useful for the assessment of the PSF because the data suffer from 
"pile-up"$\sp{13,14}$, i.e. there is more than one count per pixel per read-out 
frame, which 
happens predominantly in the center of the image with the intensity 
underestimated there. In "small window mode" the observational field is reduced 
and the CCD read-out time is shorter such that the effect of "pile-up" is 
reduced accordingly. For this mode the PSF analyis could be done 
for the last two sources mentioned above. But the radial extent 
of the PSF is limited accordingly. The PSFs 
taken on ground can therefore only be compared with the in-orbit 
results out to the angular distance imposed by the "small window" 
size. The wings of the PSFs are not accessible with these early 
measurements. Likewise, the encircled energy functions,  
i.e. the radially integrated PSFs, have been normalized to the value 
measured on ground 
at the angular distance corresponding to the 
"small window" size.

A major difference between the conditions on ground and in orbit 
is the type of detector used in the focal plane, in particular its 
pixel size. The CCD used on ground had a pixel size of 27 $\mu$m equivalent 
to 0.65 arcsec. The CCDs used in orbit have a pixel size of 40 $\mu$m for 
the MOSs and 150 $\mu$m for the pn, which relates to 1.1 arcsec and 
4.1 arcsec, respectively. In particular the pn-CCD pixel size appears too 
large to resolve the core of the PSF. 

The in-orbit imaging performance of the telescopes is presented 
in terms of contour lines, describing the surface brightness across 
the image 
(Figure 1), the PSF  
 and the encircled energy function (Figure 2) 
for FM3/MOS1 at 1.5 keV. Figure 3 and Figure 4 show the corresponding data for FM4/MOS2 
at 1.5 keV as well.
The in-orbit results are compared with the equivalent set of data measured 
at PANTER. The agreement between on ground and in-orbit 
measurements is striking. It is noted that the contour line 
plots have been produced using a smoothing algorithm including the 
adjacent three pixels in either direction, which tends 
to increase the width of the core of the PSF. 
For the determination of the FWHM of the PSF the unsmoothed contours have 
been used giving the FWHM values listed in Table~\ref{tab:perform}. 
Figure 5 shows 
the encircled energy function for FM2/pn at 1.5 keV for  
the source PSR 0540-69 and  the 
encircled energy 
function for LMC X-3 for  an energy band of (8 $\pm 1$) keV. 
The corresponding ground calibration measurements are not shown but the results 
 are identical within very small excursions  with the in-orbit 
performance. 
Because of the comparatively large pixel size of the pn-CCD details of the core 
of the PSF are not accessible in orbit and the FWHM of the PSF of FM2 cannot 
be determined. The PSF of FM2 is undersampled by the pn-CCD.

\begin{figure}
\begin{center}
\begin{tabular}{c}
\hbox{\psfig{figure=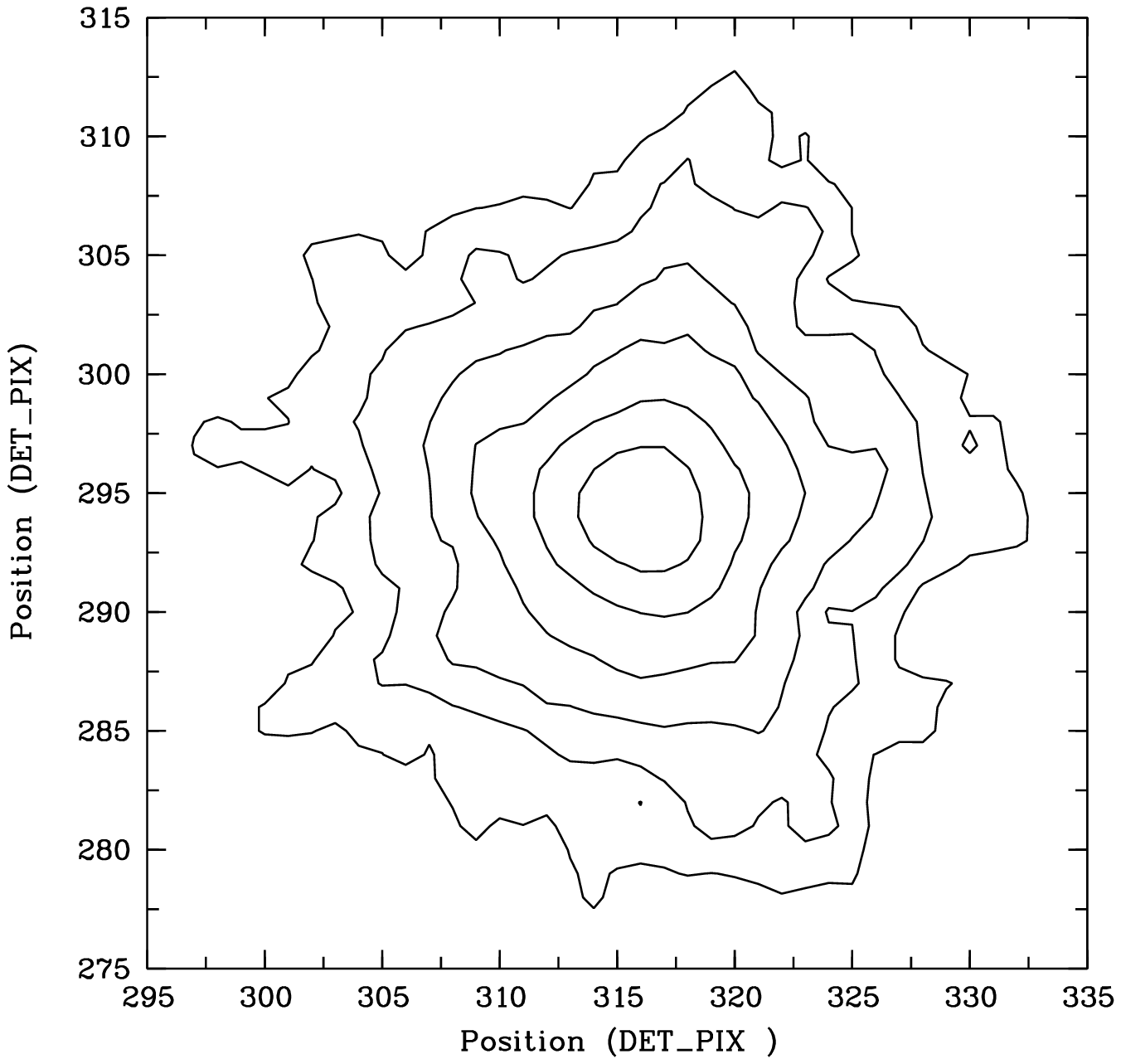,angle=0,%
bbllx=105pt,bblly=230pt,bburx=500pt,bbury=605pt,width=8.0cm,clip=}
\hfil
\psfig{figure=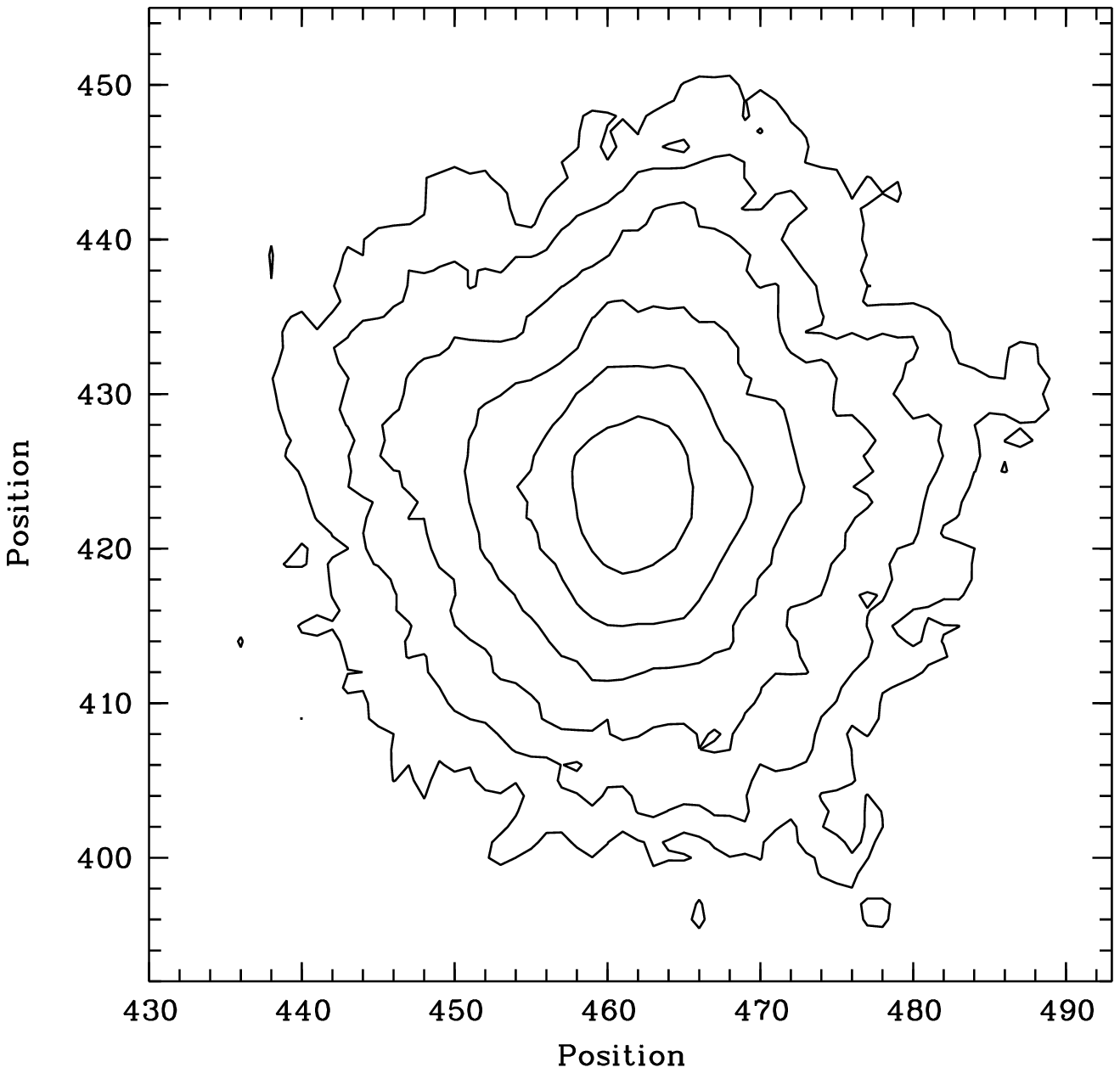,angle=0,%
bbllx=105pt,bblly=230pt,bburx=500pt,bbury=605pt,width=8.0cm,clip=}}
\end{tabular}
\end{center}
\caption {Surface brightness of a point source image taken with mirror module FM3 in orbit 
({\it left}) and on ground 
({\it right}). The size of the frame is 44" by 44" for both the in-orbit and on-ground data.
 Contours are stepped by a factor of two in brightness. In-orbit contours appear to be smoother 
because of much lower counting statistics.}
\end{figure}

\begin{figure}
\begin{center}
\begin{tabular}{c}
\hbox{\psfig{figure=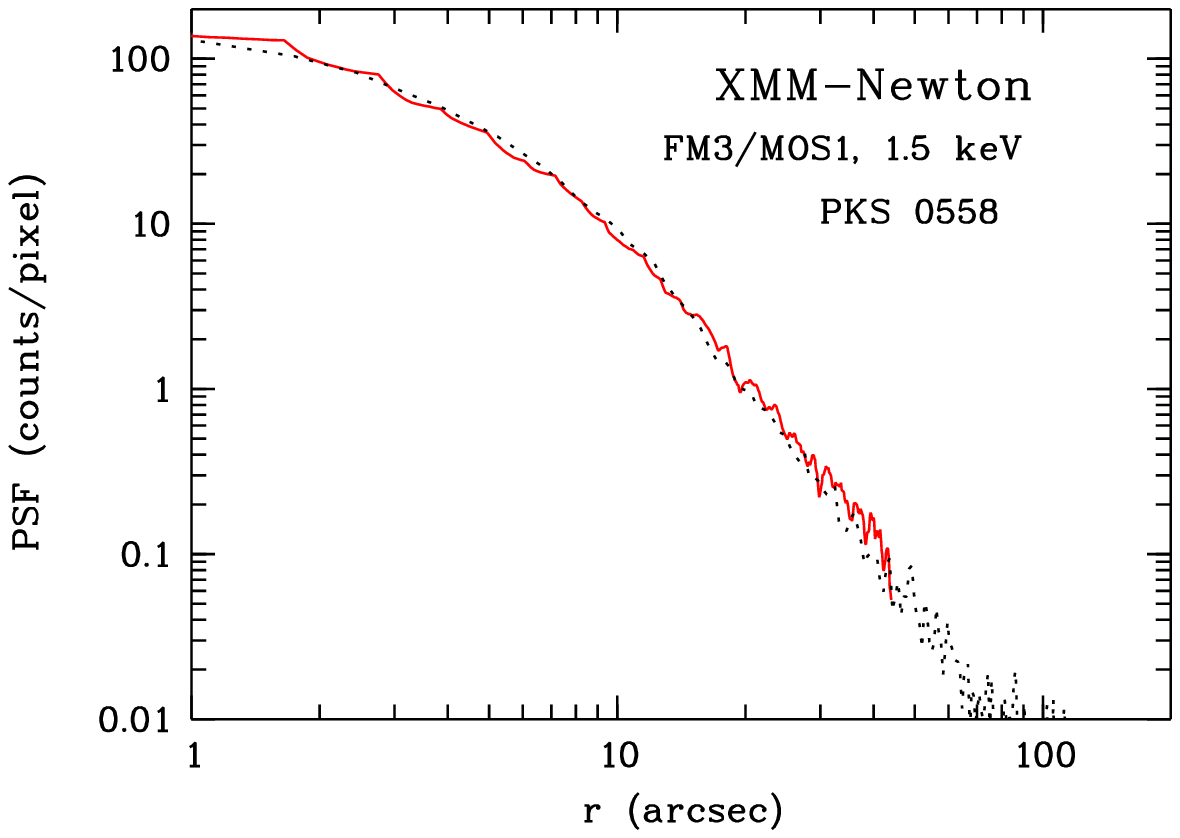,angle=0,%
bbllx=100pt,bblly=355pt,bburx=450pt,bbury=600pt,width=8.0cm,clip=}
\hfil
\psfig{figure=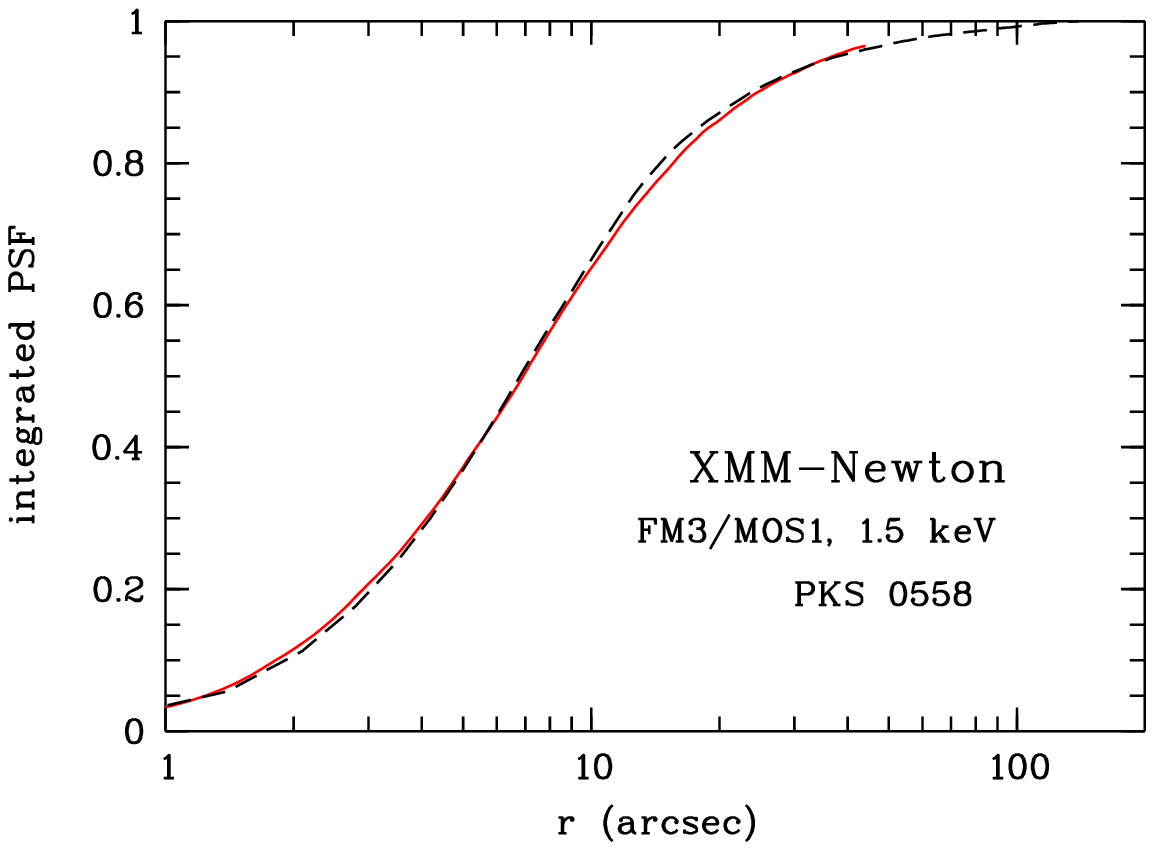,angle=0,%
bbllx=105pt,bblly=355pt,bburx=450pt,bbury=600pt,width=8.0cm,clip=}}
\end{tabular}
\end{center}
\caption {Point spread function ({\it left}) and  
encircled energy function ({\it right})  
of mirror module FM3 (solid line: in orbit, 
dashed line: on ground).}
\end{figure}

\begin{figure}
\begin{center}
\begin{tabular}{c}
\hbox{\psfig{figure=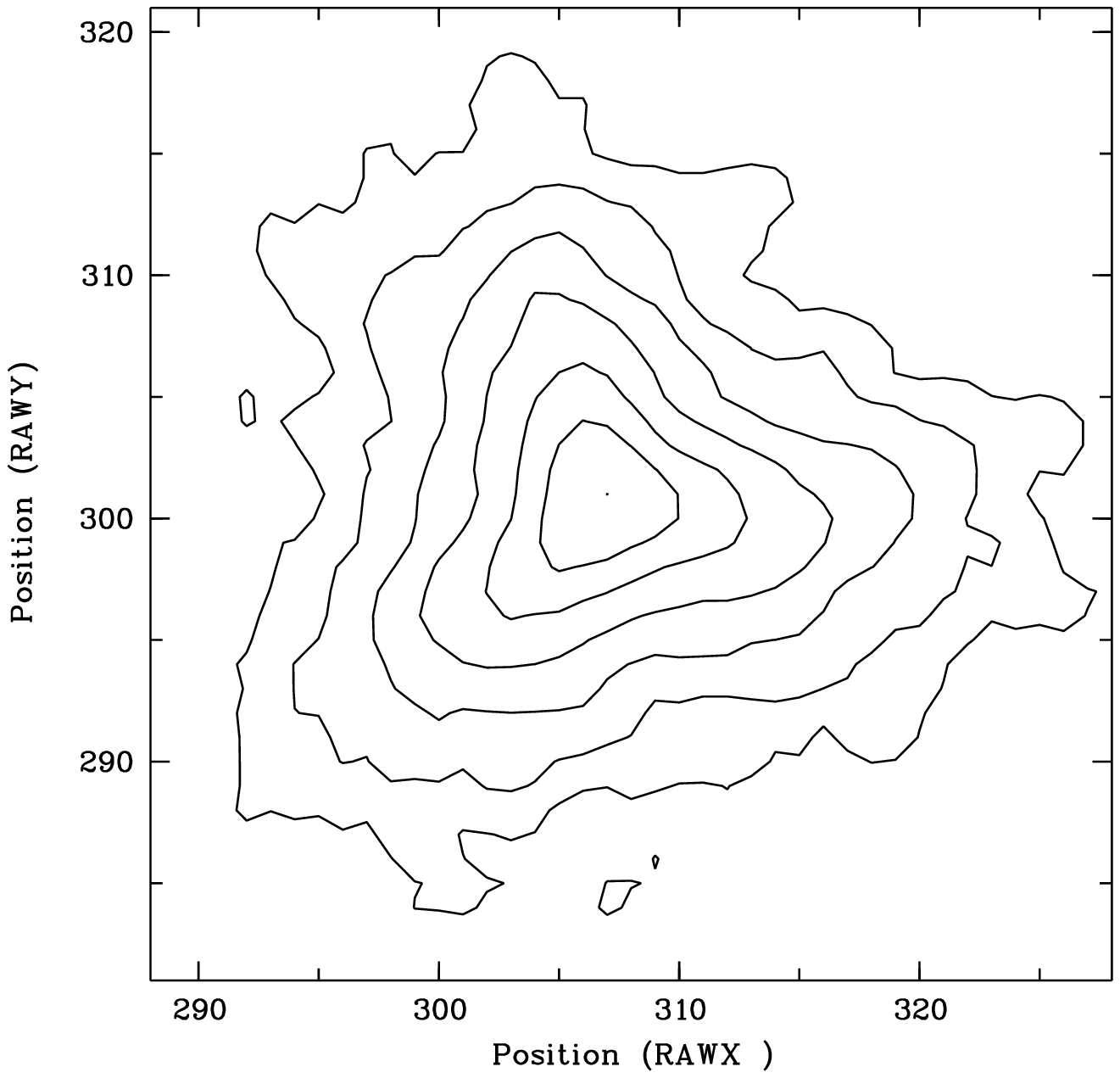,angle=0,%
bbllx=105pt,bblly=230pt,bburx=500pt,bbury=605pt,width=8.0cm,clip=}
\hfil
\psfig{figure=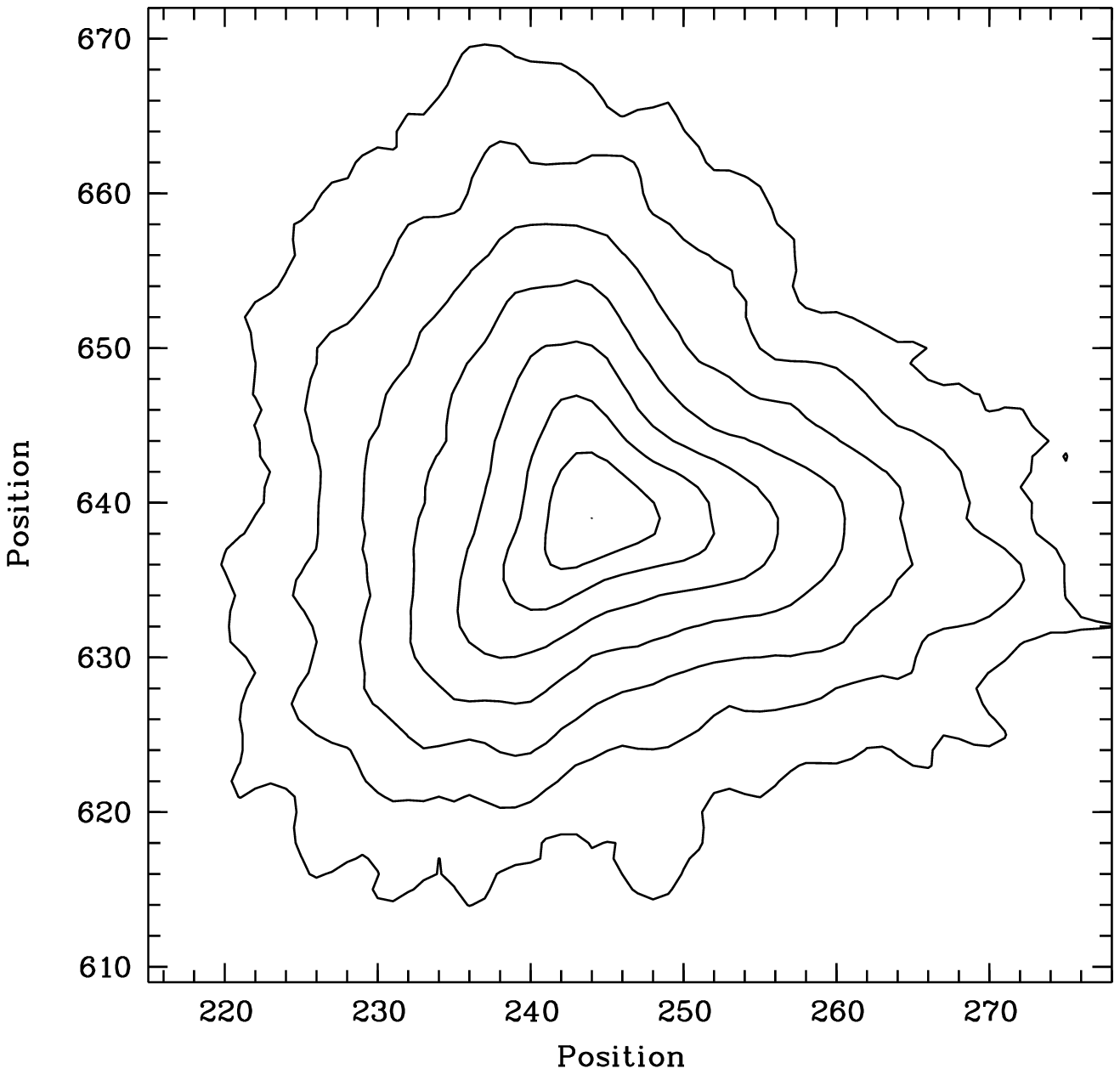,angle=0,%
bbllx=105pt,bblly=230pt,bburx=500pt,bbury=605pt,width=8.0cm,clip=}}
\end{tabular}
\end{center}
\caption {Surface brightness of a point source image taken with mirror module FM4 in orbit
({\it left}) and on ground
({\it right}). The size of the frame is 44" by 44" for both the in-orbit and on-ground data. 
Contours are stepped by a factor of two in brightness. In-orbit contours appear to be smoother
because of much lower counting statistics.}
\end{figure}

\begin{figure}
\begin{center}
\begin{tabular}{c}
\hbox{\psfig{figure=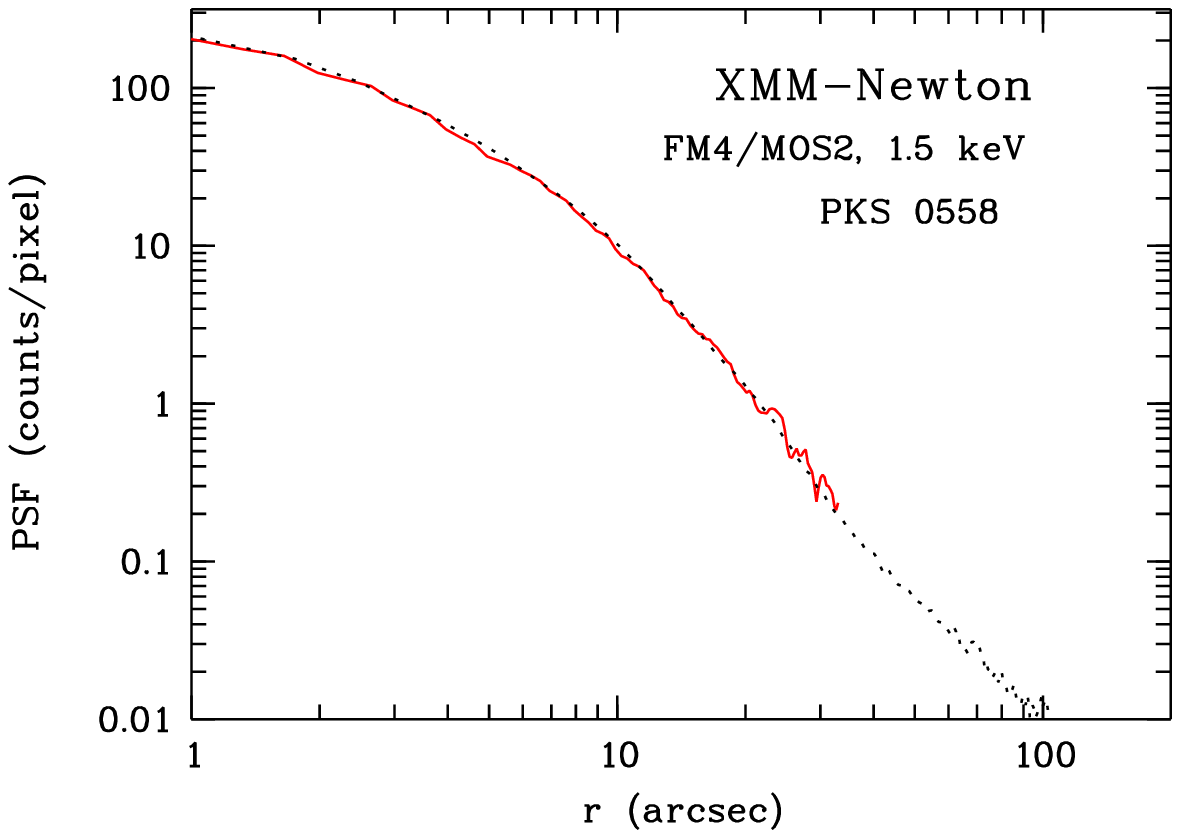,angle=0,%
bbllx=100pt,bblly=355pt,bburx=450pt,bbury=600pt,width=8.0cm,clip=}
\hfil
\psfig{figure=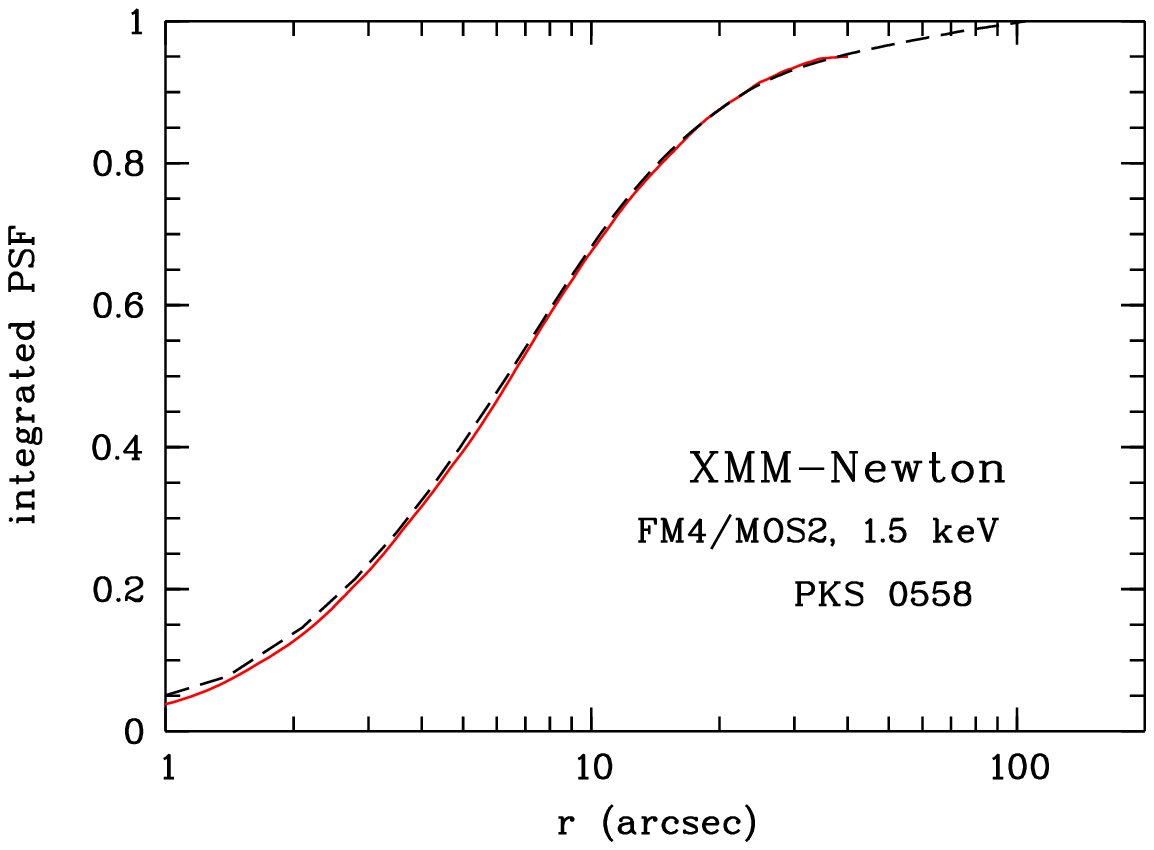,angle=0,%
bbllx=100pt,bblly=355pt,bburx=450pt,bbury=600pt,width=8.0cm,clip=}}
\end{tabular}
\end{center}
\caption {Point spread function ({\it left}) and
encircled energy function ({\it right})
of mirror module FM4 (solid line: in orbit,
dashed line: on ground).}
\end{figure}

\begin{figure}
\begin{center}
\begin{tabular}{c}
\hbox{\psfig{figure=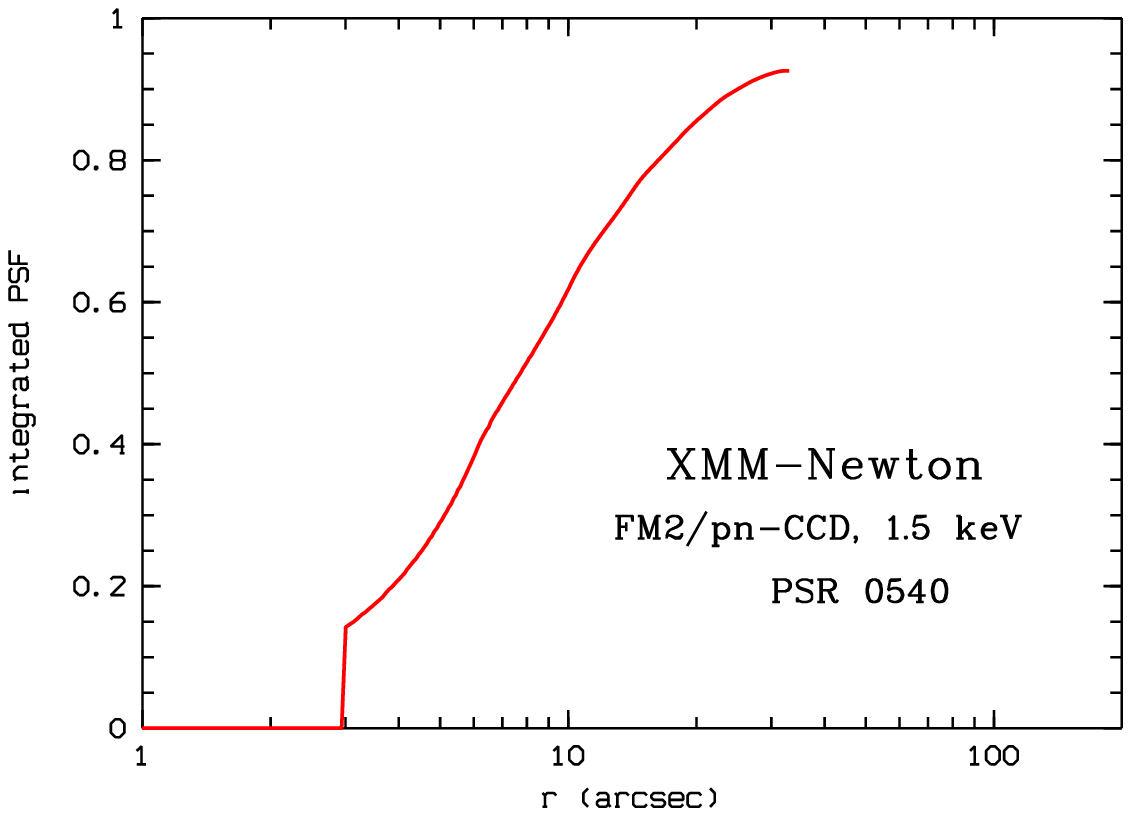,angle=0,%
bbllx=115pt,bblly=365pt,bburx=450pt,bbury=600pt,width=8.0cm,clip=}%
\hfil
\psfig{figure=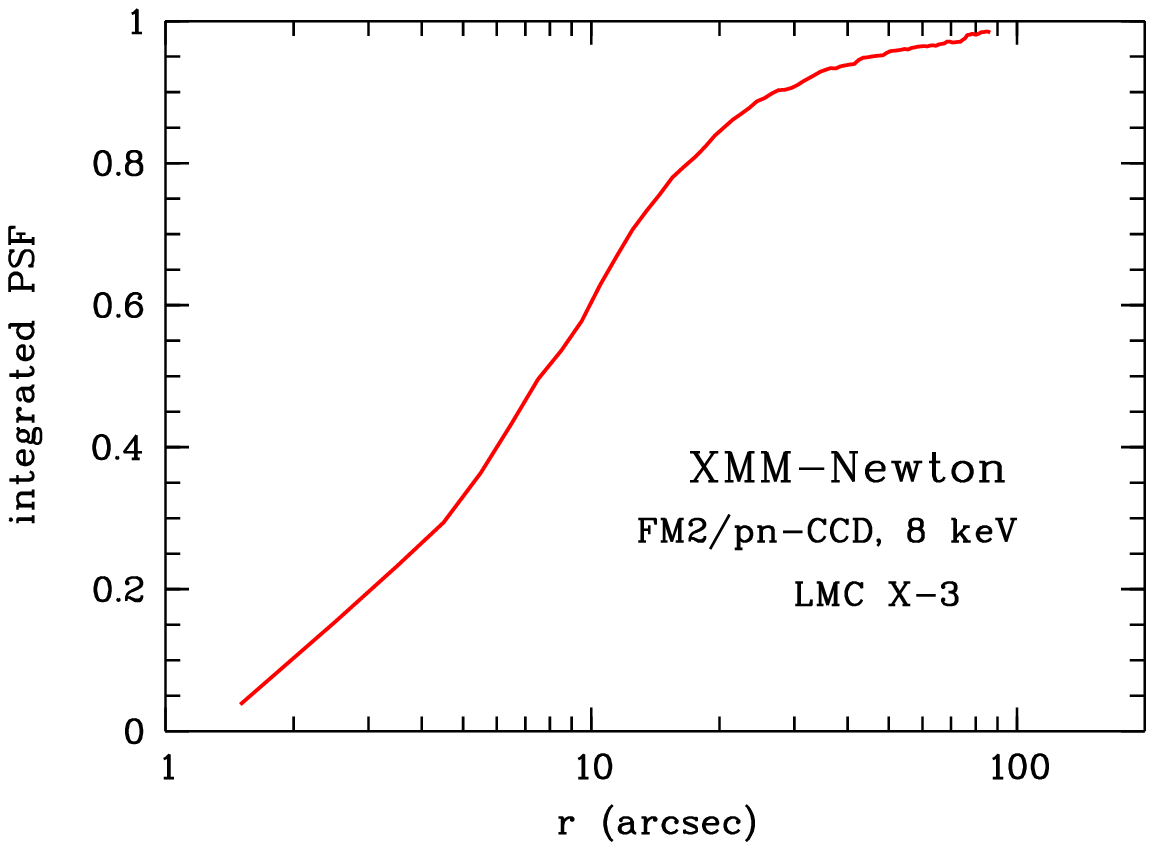,angle=0,%
bbllx=110pt,bblly=360pt,bburx=450pt,bbury=600pt,width=8.0cm,clip=}}
\end{tabular}
\end{center}
\caption {Encircled energy function at 1.5 keV ({\it left}) and
at 8 keV ({\it right})
of mirror module FM2 in orbit.}
\end{figure}

Table~\ref{tab:perform} summarizes the results in terms of characteristic  
numbers like 
FHMW and HEW (half energy width of the encircled energy function).

\begin{table} [h]  
\caption{Summary of the imaging performance of the XMM-Newton 
X-ray telescopes.}
\label{tab:perform}
\begin{center}      
\begin{tabular}{|l|r|r|r|r|r|r|r|r|r|} 
\hline
\rule[-1ex]{0pt}{3.5ex}   & 1.5 keV & 1.5 keV & 1.5 keV  & 1.5 keV & & 8 keV & 8 keV & 8 keV & 8 keV\\
\hline
\rule[-1ex]{0pt}{3.5ex}   & orbit  & ground  & orbit  & ground & &orbit  & ground & orbit  & ground  \\
\rule[-1ex]{0pt}{3.5ex}  telescope & FWHM  & FWHM  & HEW  & HEW & &FWHM  & FWHM & HEW  & HEW  \\
\rule[-1ex]{0pt}{3.5ex}   &  (") &  (") &  (") &  (") & & (") & (") & (") &  (") \\
\hline
\rule[-1ex]{0pt}{3.5ex}  FM2/pn & $<$ 12.5 & 6.6 & 15.2 & 15.1 & & $<$12.5 & 6.6 & 15.0 & 14.8   \\
\rule[-1ex]{0pt}{3.5ex}  FM3/MOS1 &  4.3 & 6.0 & 13.8 & 13.6 & & -  & 5.1 & - & 12.5  \\
\rule[-1ex]{0pt}{3.5ex}  FM4/MOS2 &  4.4 & 4.5 & 13.0 & 12.8 & & -  & 4.2 & - & 12.2  \\
\hline 
\end{tabular}
\end{center}
\end{table}

\section{CONCLUSIONS}

Providing that the PSFs of the three X-ray telescopes on board of 
XMM-Newton have not developed significantly bright wings outside 
of about 1 arcmin radius from ground to early orbit the following 
conclusions can be drawn.

\begin{itemize}
\item{
From the contour lines and the FWHM of the PSFs and the agreement 
with the ground calibration measurements it follows 
that all three focal plane cameras are in focus.} 
\item{
The PSFs in orbit are basically identical with those measured on ground;
the core of FM2 cannot be resolved in orbit because of the 
large pn-CCD pixel size.} 
\item{
The encircled energy functions of FM2, FM3 and 
FM4 measured  in orbit at 1.5 keV agree with those measured on ground, at least 
out to an angular radius of about 40 arcsec.
At 8 keV the FM2 shows the same encircled energy function in orbit
 as measured on ground.}
\end{itemize}

Beyond this direct comparison further conclusions can be drawn which 
are related to the mechanical integrity of the telescopes and the 
X-ray calibration test set-up.

\begin{itemize}
\item{
The X-ray telescopes have not changed their shape during storage on ground 
or by launch loads.}
\item{
The focal plane instrumentation was correctly installed at the required 
position 
and no significant movement occurred during launch.}
\item
{The mounting and fixture of the mirrors in just one spider support
 structure at the parabolas' front is sufficient to maintain 
the mirrors' performance.}
\item
{Effects of gravity which would potentially distort the mirrors during 
horizontal X-ray testing in the MPE PANTER test facility are not 
apparent concluded from the comparison 
of the ground measurements with the in-orbit results of the PSF.}
\item
{The divergence of the X-ray beam in the PANTER test geometry 
does not limit the assessment of the PSF. Interestingly the non-illumination 
of the first third of the parabolas' sections does not change 
the PSF of the XMM-Newton telescopes. This 
means that the angular distortions along the meridian lines of the parabola 
mirrors imposed by the mounting at their front ends are not significantly 
larger than along the remainder of the meridian lines towards the 
hyperbolas' exits.}
\end{itemize}

\acknowledgements

The provision of the pn-CCD  and MOS-CCD in orbit data of XMM-Newton
 by the EPIC team led by the Principal Investigator Martin Turner 
of the Leicester University,  UK is acknowledged.

\section*{REFERENCES}

\begin{enumerate}
\item "A Cosmic X-ray Spectroscopy Mission", Proc. of a Workshop 
      held in Lyngby, Denmark on 24-26 June 1985, {\it{Proc. ESA}} {\bf{SP-239}}, 1985.
\item B. Aschenbach, O. Citterio, J.M. Ellwood, P. Jensen, P. de Korte, A. Peacock and 
      R. Willingale, "The High-Throughput X-ray Spectroscopy Mission", Report of the 
      Telescope Working Group, {\it{Proc. ESA}} {\bf{SP-1084}}, 1987.
\item B. Aschenbach and Heinrich Br\"auninger, " Grazing Incidence Telescopes for ESA's 
      X-ray astronomy mission XMM", in {\it{X-ray Instrumentation in Astronomy II}}, 
      L. Golub, ed., {\it{Proc. SPIE}} {\bf{982}}, pp. 10-15, 1988.
\item B. Aschenbach, "Re-Design of the XMM Optics", {\it{Techn. Note XMM/O/MPE}}, 1987.
\item B. Aschenbach, H. Br\"auninger, K.-H. Stephan and Joachim Tr\"umper, 
      "X-ray test facilities at Max-Planck-Institut Garching", in 
      {\it{Space Optics - Imaging X-ray Optics Workshop}}, M. Weisskopf, ed., 
      {\it{Proc. SPIE}} {\bf{184}}, pp. 234-238, 1979.
\item D. de Chambure, R. Lain\`e, K. van Katwijk, A. Valenzuela, G. Grisoni {\it{et al.}}, 
      "Lessons learnt from the development of the XMM optics", Intern. Symposium on the 
      Optical Design and Production, Berlin 1999, {\it{EUROPTO Series 1999}}, in press.
\item B. Aschenbach, "Design of an X-Ray Baffle-System", 
      {\it{Techn. Note XMM-TS-FMP004}}, 1996.
\item D. de Chambure, R. Lain\`e, K. van Katwijk, W. R\"uhe, D. Schink {\it{et al.}},
      "The X-ray Baffle of the XMM telescope: Development and Results", Intern. Symposium on the
      Optical Design and Production, Berlin 1999, {\it{EUROPTO Series 1999}}, paper no. 
      3737-53, in press.
\item E. Pfeffermann, U.G. Briel, H. Hippmann, G. Kettenring, G. Metzner {\it{et al.}},
      "The focal plane instrumentation of the ROSAT telescope", 
      {\it{Proc. SPIE}} {\bf{733}}, pp. 519-532, 1986.  
\item R. Egger, B. Aschenbach, H. Br\"auninger, W. Burkert, T. D\"ohring and A. Oppitz,
      "X-ray calibration of the mirror module FM2 at PANTER test facility - 
      test results", {\it{Techn. Note XMM-TS-PA063/970829}}, 1997.
\item R. Egger, B. Aschenbach, H. Br\"auninger, W. Burkert, T. D\"ohring and A. Oppitz,
      "X-ray calibration of the mirror module FM3 at PANTER test facility -
      test results", {\it{Techn. Note XMM-TS-PA070/980115}}, 1998.
\item R. Egger, B. Aschenbach, H. Br\"auninger, W. Burkert, K. Molthagen and A. Oppitz,
      "X-ray calibration of the mirror module FM4 at PANTER test facility -
      test results", {\it{Techn. Note XMM-TS-PA077/980723}}, 1998.
\item U.G. Briel, B. Aschenbach, M. Balasini, H. Br\"auninger, W. Burkert, K. Dennerl 
      {\it{et al.}},
      "The in-orbit Performance of the EPIC-PN CCD-Camera on Bord of XMM-Newton", 
      in {\it{X-ray Optics, Instruments and Missions}},
      J. Tr\"umper and B. Aschenbach, eds., {\it{Proc. SPIE}} {\bf{this volume}}, 2000.
\item L. Str\"uder, N. Meidinger, E. Pfefferman, R. Hartmann, H. Br\"auninger
      {\it{et al.}},
      "X-ray pn-CCDs on the XMM-Newton Observatory",
      in {\it{X-ray Optics, Instruments and Missions}},
      J. Tr\"umper and B. Aschenbach, eds., {\it{Proc. SPIE}} {\bf{this volume}}, 2000. 
\item J.W. den Herder, A.J. den Boggende, G. Branduardi-Raymont, A.C. Brinkman, J. 
      Cottam 
      {\it{et al.}},
      "Description and performance of the reflection grating spectrometer on board of 
      XMM-Newton",
      in {\it{X-ray Optics, Instruments and Missions}},
      J. Tr\"umper and B. Aschenbach, eds., {\it{Proc. SPIE}} {\bf{this volume}}, 2000.
\end{enumerate}
\end{document}